\documentclass[prb,aps,nobalancelastpage,amssymb,twocolumn,citeautoscript]{revtex4}

\usepackage{graphicx}

\begin{document}

\title{Synthesis, characterization and physical properties of layered bismuthide PtBi$_2$}

\author{ C. Q. Xu$^1$, X. Z. Xing$^2$, Xiaofeng Xu$^{1}$\footnote[2]{Corresponding author: xiaofeng.xu@hznu.edu.cn}, L. Q. Che$^3$, Bin Chen$^4$, Xin Lu$^3$, Jianhui Dai$^1$, Z. X. Shi$^2$}

\affiliation{$^1$Department of Physics and Hangzhou Key Laboratory of Quantum Matters, Hangzhou Normal University, Hangzhou 310036, China}
\affiliation{$^2$Department of Physics and Key Laboratory of MEMS of the Ministry of Education, Southeast University, Nanjing 211189, China}
\affiliation{$^3$Center for Correlated Matter and Department of Physics, Zhejiang University, Hangzhou, 310058, China}
\affiliation{$^4$Department of
Physics, University of Shanghai for Science $\&$ Tehcnology , Shanghai, China}

\date{\today}

\begin{abstract}
We report details of single crystal growth of stoichiometric bismuthide PtBi$_2$ whose structure consists of alternate stacking of Pt layer
sandwiched by Bi bilayer along the $c$-axis. The compound crystallizes in space group P-3 with a hexagonal unit cell of $a$=$b$=6.553$\AA$,
$c$=6.165$\AA$. The magnetization data show opposite sign for fields parallel and perpendicular to the Pt layers, respectively. The $T$-dependent
resistivity is typical of a metal and the magnetic response shows clear two types of charge carriers and the validity of the semi-classical Kohler's
rule. Its physical properties was discussed in comparison with recently proposed topological superconductor $\beta$-PdBi$_2$.
\end{abstract}

\maketitle
\section{Introduction}
\label{Intro}

Binary bismuthide $\beta$-PdBi$_2$ has attracted much interest recently as a promising candidate of topological superconductor
(TS)\cite{Imai,chu15,Herrera,Sakano15,Kacmarcik,LuXin16}. Topological superconductivity is a new state of matter possessing symmetry-protected
surface states while the bulk states are fully gapped by superconducting pairing\cite{Hasan10,Cava10,FuPRB,FuPRL}. The Majorana fermions are believed
to exist on the surface or vortex core in such TSs, which may not only be of scientific importance, but also can lead to a wide-ranging applications
in microelectronic devices and quantum computing. The centrosymmetric stoichiometric $\beta$-PdBi$_2$ ($T_c$$\sim$ 5 K) was claimed to be
topologically nontrivial in view of the observation of the topologically-protected surface modes by spin- and angle-resolved ARPES\cite{Sakano15}.
However, no Andreev bound states associated with Majorana fermions are detectable through point-contact spectroscopy\cite{LuXin16}, in sharp contrast
to the cases in Cu-intercalated Bi$_2$Se$_3$\cite{Ando11,Ando11PCS} and In-doped SnTe\cite{Ando13}. On the other hand, it becomes the common wisdom
that spin-orbit interaction (SOI) in heavy elements is crucial for the topological states. It is therefore heuristic to ask what if we replace Pd by
heavier Pt element with enhanced SOI.

\begin{figure}
\includegraphics[width=9cm,keepaspectratio=true]{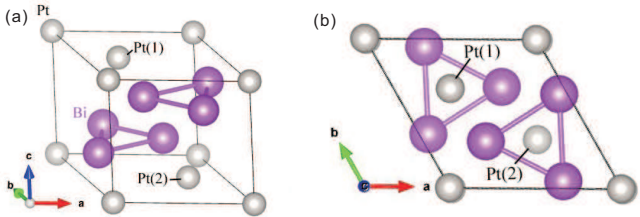}
\caption{(Color online) Crystal lattice of PtBi$_2$. (a) The primitive unit cell for hexagonal PtBi$_2$. The coordinates of Pt(1) and Pt(2) are
($\frac{a}{3},\frac{2a}{3}, 0.92c)$ and ($\frac{2a}{3},\frac{a}{3}, 0.08c)$, respectively. (b) The structure as seen from a perspective along the
$c$-axis. } \label{Fig1}
\end{figure}

In this study, we substituted Pt for Pd in PdBi$_2$ and found that this new material actually crystallizes in a distinct structure. Unlike
$\beta$-PdBi$_2$ which has the tetragonal structure in an $I$4/$mmm$ space group, PtBi$_2$ crystallizes in space group P-3 with a hexagonal unit
cell of $a$=$b$=6.553$\AA$, $c$=6.165$\AA$\cite{book-PtBi2}. It is also different from its homologue PtBi superconductor ($T_c$=1 K) with a
monoclinic unit cell\cite{Matthias}. The in-plane resistivity of PtBi$_2$ shows metallic behaviors down to 2 K, the lowest temperature studied in
this work. The intra-plane and inter-plane magnetization displays pronounced anisotropy, being diamagnetic with field aligned along the plane and
paramagnetic when field is perpendicular to the plane. The magnetoresistance (MR) and Hall resistivity measured on the same sample both show two
types of carriers and the former one scales well to the semi-classical Kohler's rule\cite{Kohler1938,NieLuo02}.

\section{Experimental}
\label{Exp}

PtBi$_2$ single crystals were fabricated via a melt-growth method. The starting materials of high purity, Bi powder(4N) and Pt powder (4N), were
mixed thoroughly in the prescribed molar ratio of Bi:Pt = 2:1 (2 g in total weight). All these preparations were performed in a glove box filled with
protective argon gas (both H$_2$O and O$_2$ contents were limited below 0.1ppm). The mixtures were loaded and sealed in an evacuated quartz tube.
This quartz tube was then heated to 700$^\circ$C quickly in a sintering furnace and kept at this temperature for 48h, before being slowly cooled down
to 450$^\circ$C(3$^\circ$C/h), and finally being quenched into cold water. Large pieces of dark-gray plate-like PtBi$_2$ single crystals of typical
7-8 mm in length were harvested.

Energy dispersive x-ray (EDX) spectrometry confirms the stoichiometric ratio of the chemical composition (32.8 : 67.2 $\pm$ 3.0\% in molar percentage
for Pt:Bi). The structure of crystals was characterized by powder x-ray diffraction (XRD) at room temperature using a Rigaku diffractometer with Cu
$K$$\alpha$ radiation and a graphite monochromator. Lattice parameters were obtained by Rietveld refinements. The magnetization was measured by
vibrating sample magnetometry using a Quantum Design MPMS-5 system. Measurements of MR and Hall effect were performed on the same sample by changing
the field polarities. Signal even in field was defined as MR and the odd component was calculated as Hall resistivity.

\begin{figure}
\includegraphics[width=9cm,keepaspectratio=true]{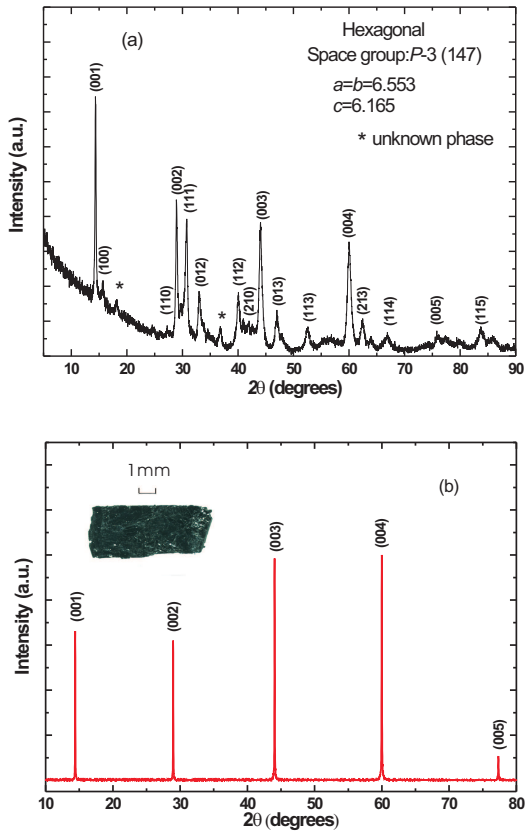}
\caption{(Color online) Panels (a) and (b) represent the powder XRD patterns and single crystal XRD diffraction peaks, respectively. The asterisks in
panel (a) mark the possible impurity phases. The optical image of a single-crystal sample is shown in the inset of panel (b).} \label{Fig2}
\end{figure}

\begin{figure}
\includegraphics[width=9.2cm,keepaspectratio=true]{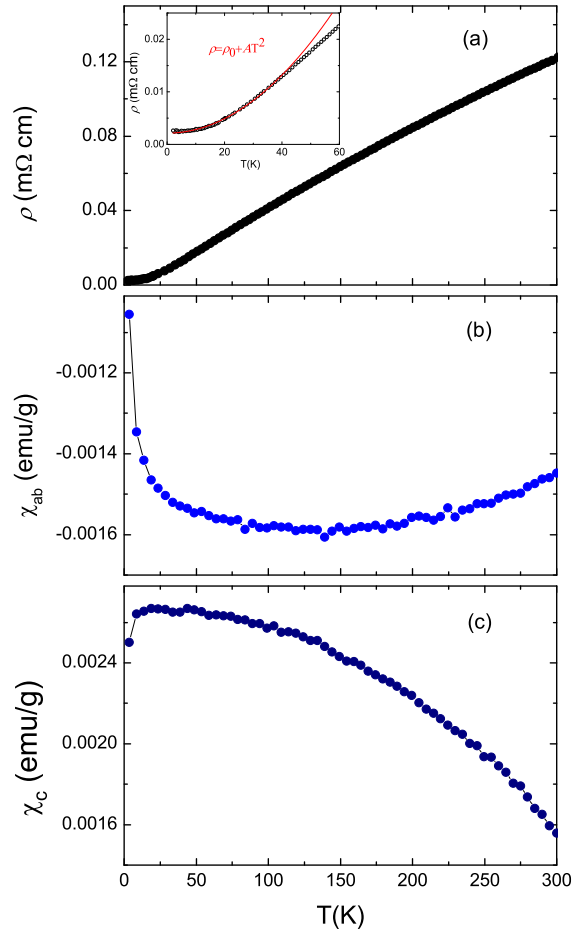}
\caption{(Color online) (a) Zero-field resistivity curve down to 2K. The low-$T$ resistivity fits to $\rho=\rho_0+AT^2$ very well below $\sim$35K
(see the upper-left inset). (b) and (c) show the in-plane and inter-plane susceptibility under a field of 5koe, respectively.} \label{Fig3}
\end{figure}

\section{Results}
\label{Results}

The schematic view of the crystal structure of PtBi$_2$ is shown in Fig. 1. It crystallizes in a hexagonal structure with the space group P-3
(No.147). Its structure consists of alternate stacking of 2D Pt layers and bismuth bilayers along the $c$-axis. In one primitive unit cell, there are
three Pt atoms, one being located at the corner of the polyhedron and the other two labelled as Pt(1) and Pt(2) in Fig. 1. The Bi atoms are
trigonally-coordinated. The XRD pattern of PtBi$_2$ crystal is presented in Fig.2. A small trace of impurity phase, marked by the asterisks in panel
(a), was detectable in the powder X-ray pattern and only (00$\ell$) diffraction peaks were observed in the single-crystal X-ray, indicating good
$c$-axis orientation of the as-grown samples. The calculated lattice parameters are $a$=$b$=6.553$\AA$, $c$=6.165$\AA$, in consistence with previous
reported results\cite{book-PtBi2}.

\begin{figure*}
\includegraphics[width=18cm,keepaspectratio=true]{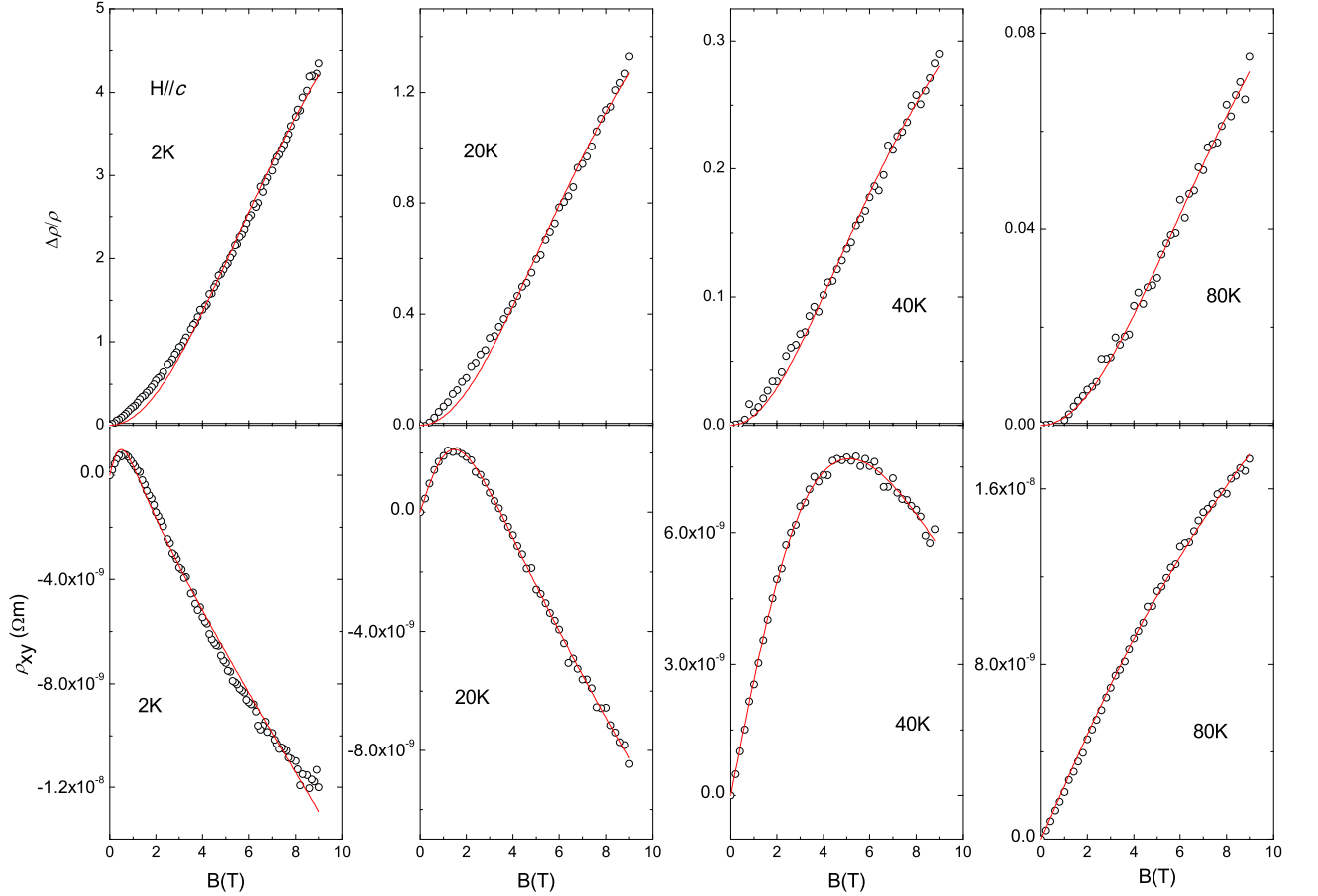}
\caption{(Color online) The magnetoresistance (upper panel) and Hall resistivity (lower panel), both measured on the same crystal with the same
electrical contacts, at several selected temperatures. The red solid curves delineate the fits to two-band carrier model.} \label{Fig4}
\end{figure*}

Zero-field in-plane resistivity is plotted in Fig. 3. The room temperature resistivity is about 0.12 m$\Omega$cm and it is metallic down to the
lowest temperature we measured (2K). The residual resistivity ratio is approximately 50 for our samples, indicative of good sample quality. The
sample is better characterized by the susceptibility measurements thereafter. Remarkably, the magnetization of the sample shows large anisotropy with
respect to the field orientations. As illustrated in Fig. 3, the in-plane magnetization $\chi_{ab}$ is diamagnetic and varies little with $T$ down to
20K, below which it displays a significant upturn, whereas the inter-plane $\chi_c$ is paramagnetic instead and increases linearly with decreasing
$T$, followed by a downward trend below 20K. The origin of these intriguing magnetization behaviors is not clear.

The magnetoresistive and Hall response of a material can open a avenue for exploring the dispersion and dynamics of the charge carriers. First, in
PtBi$_2$, it is noted that the absolute value of the MR, defined as $\frac{\Delta\rho}{\rho}$, is rather large, reaching $>$400\% at 2K in a magnetic
field of 9T. This large MR implies a rather large electron mean free path, hence a long relaxation time. However, this MR is damped very fast with
increasing $T$, as seen from the upper panels of Fig. 4. Second, in single-band metals, the MR at small fields is usually quadratic and the Hall
resistivity varies linearly with field. However, in the two-band Drude model, on the assumption of the field-independent carrier density and
relaxation time, $\frac{\Delta\rho}{\rho}(H)$ and $\rho_{xy}(H)$ can be written as\cite{Greene07,Rullier-Albenque09,Hussey10,Rullier-Albenque12}

\begin{eqnarray}
\frac{\Delta\rho}{\rho}=\frac{\sigma_h\sigma_e(\sigma_hR_h-\sigma_eR_e)^2H^2}{(\sigma_h+\sigma_e)^2+\sigma_h^2\sigma_e^2(R_h+R_e)^2H^2}\label{eqn:one}
\end{eqnarray}

\begin{eqnarray}
\rho_{xy}(H)=\frac{\sigma_h^2R_h+\sigma_e^2R_e+\sigma_h^2\sigma_e^2R_hR_e(R_h+R_e)H^2}{(\sigma_h+\sigma_e)^2+\sigma_h^2\sigma_e^2(R_h+R_e)^2H^2}H\label{eqn:two}
\end{eqnarray}

\noindent where $\sigma_e(h)$ and $R_e(h)$ are electrical conductivity and Hall coefficient for electron (hole) band, respectively. The MR and the
Hall signal for PtBi$_2$ sample are exemplified in Fig 4 at some selected temperatures. Although the individual curves can be fitted with the above
two-band equations reasonably well, plotted as the red solid line in each panel, we failed to model these two transport coefficients simultaneously
with the same set of four parameters. These difficulties may arise from the simple assumption of the field independent charge carrier density and
scattering, in analogy to the case in cuprates\cite{Greene07}. Nevertheless, given the quality of our fitting and the strong non-linearity of the
Hall resistivity, we strongly believe that the transport properties of this compound are governed by two-band charge carriers.

In standard metals, the MR $\Delta \rho$/$\rho$ at a certain temperature under a field $H$ has a general form known as the Kohler's
rule\cite{Kohler1938,NieLuo02}: $\Delta \rho$/$\rho$=$f$($H/\rho$). This rule can be derived from Boltzmann transport theory, on the assumption of
constant carrier number with $T$ and a single scattering rate on the Fermi surface. From this rule, $\Delta \rho$/$\rho$ is literally independent of
$T$ such that the plots of $\Delta \rho/\rho_0$ as a function of $H/\rho$ at distinct temperatures will collapse onto a single curve. Interestingly,
this rule, albeit its semiclassical origin, was found to be well obeyed in a large number of metals from conventional metals to some quantum matters.
These involve the metals with two types of carriers\cite{NieLuo02}, the pseudogap phase of the underdoped cuprates\cite{Greven14},
quasi-one-dimensional metals\cite{Narduzzo07,Xu15} as well as some topological semimetals\cite{Coldea}. We examined this rule in PtBi$_2$ (Fig. 5)
and found that it is well obeyed in this material, over a wide field range (up to 9T) and a broad $T$ window (2K-100K. Above 100K, the MR tends to be
negligible). Moreover, the longitudinal MR in PtBi$_2$ (H $\|$ I $\|$ $ab$) also shows two types of charge carriers and the validity of the Kohler's
rule (data not shown).

\begin{figure}
\includegraphics[width=9cm,keepaspectratio=true]{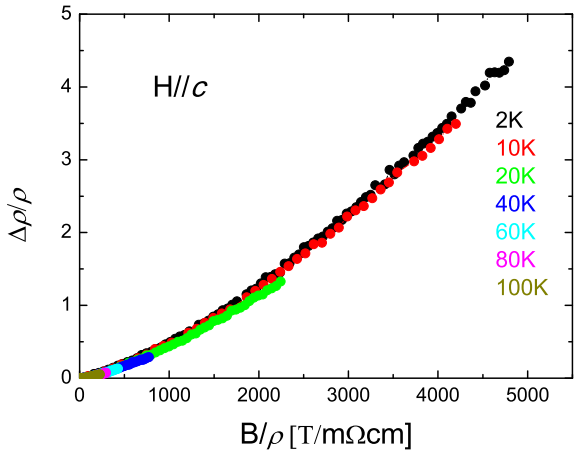}
\caption{(Color online) The Kohler's plot for the MR data from Fig. 4. } \label{Fig5}
\end{figure}

\section{Discussions and Conclusion}
\label{Discussions}

In recent work by Sakano \textit{et al.}, several topologically-protected surface states were observed by spin-resolved ARPES in the TS candidate
$\beta$-PdBi$_2$\cite{Sakano15}. These non-trivial surface bands include one crossing the Fermi level and the other one forming the Dirac cone state
2 eV below the Fermi level. It was noted that these topological surface states are \textit{all} derived as a consequence of SOI, although their
respective microscopic details may be different. In PtBi$_2$, the SOI ought to be stronger. Owing to its good metallicity, however, the electrical
transport is \textit{overall} dominated by its bulk electrons and it looks more like a conventional good metal from transport perspective. In this
material, the possible quantum linear MR arising from the degenerate Dirac fermions in the quantum limit is not observed up to
9T\cite{Abrikosov98,Ong10}. Interestingly, this material was reported to superconduct below 150 mK\cite{PtBi2-0.15K}, $\sim$40 times lower than $T_c$
in PdBi$_2$. How the SOI changes the electronic structure of PtBi$_2$, and induces the non-trivial surface states, if any, await more investigations,
both theoretically and experimentally.

To summarize, we synthesized the single crystals of stoichiometric bismuthide PtBi$_2$ by a solid-state reaction method. The samples were carefully
characterized by combined procedures of XRD, (magneto-)transport and susceptibility measurements. This compound shows prominent two-band transport
behaviors with no clear signature from the possible surface states. However, the high-quality single crystals are now ready for prospective advanced
experiments, especially for ones with more surface sensitivity.

\begin{acknowledgments}
The authors would like to thank C. M. J. Andrew, A. F. Bangura for stimulating discussions. This work is sponsored by the National Key Basic Research
Program of China (Grant No. 2014CB648400), and by National Natural Science Foundation of China (Grant No. 11474080, U1432135, 11611140101). X.X.
would also like to acknowledge the financial support from the Distinguished Young Scientist Funds of Zhejiang Province (LR14A040001) and an open
program from Wuhan National High Magnetic Field Center (2015KF15).
\end{acknowledgments}



\end{document}